\documentclass[conference]{IEEEtran}
\IEEEoverridecommandlockouts

\usepackage{cite}
\usepackage[normalem]{ulem} % 引入 ulem 宏包
\usepackage{amsmath,amssymb,amsfonts}
\usepackage{algorithmic}
\usepackage{graphicx}
\usepackage{textcomp}
\usepackage{xcolor}
\usepackage{subcaption}
\usepackage{multirow}
\usepackage{booktabs}

\usepackage[ruled,vlined]{algorithm2e}
\usepackage{booktabs}
\usepackage{hyperref}
\hypersetup{hidelinks}  

\usepackage{enumitem}
\usepackage{tabularray}
\usepackage{tikz}
\def\BibTeX{{\rm B\kern-.05em{\sc i\kern-.025em b}\kern-.08em
    T\kern-.1667em\lower.7ex\hbox{E}\kern-.125emX}}
\begin{document}

\title{Generative Semantic Communication for Joint Image Transmission and Segmentation}
\author{\IEEEauthorblockN{Weiwen Yuan$^{\ast\dag\S}$, Jinke Ren$^{\ast\dag\S}$, Chongjie Wang$^{\ast\dag\S}$, Ruichen Zhang$^{\ddag}$, Jun Wei$^{\P}$, Dong In Kim$^{\diamond}$, Shuguang Cui$^{\dag\ast\S}$}
\thanks{Corresponding author: Jinke Ren (jinkeren@cuhk.edu.cn).}
\IEEEauthorblockA{$^*$FNii-Shenzhen, $^\dag$SSE, and $^\S$Guangdong Provincial Key Laboratory of Future Networks of Intelligence,\\ The Chinese University of Hong Kong (Shenzhen), Shenzhen, China\\
$^\ddag$College of Computing and Data Science, Nanyang Technological University, Singapore\\
$^{\P}$College of Computer Science and Software Engineering, Shenzhen University, Shenzhen, China\\
$\diamond$Department of Electrical and Computer Engineering, Sungkyunkwan University, Suwon, South Korea \\
E-mail: \{223010145, 223010039\}@link.cuhk.edu.cn; \{jinkeren, shuguangcui\}@cuhk.edu.cn; \\ ruichen.zhang@ntu.edu.sg;  weijun@szu.edu.cn; dongin@skku.edu}}
\maketitle

\begin{abstract}
Semantic communication has emerged as a promising technology for enhancing communication efficiency. However, most existing research emphasizes single-task reconstruction, neglecting model adaptability and generalization across multi-task systems. In this paper, we propose a novel generative semantic communication system that supports both image reconstruction and segmentation tasks. Our approach builds upon semantic knowledge bases (KBs) at both the transmitter and receiver, with each semantic KB comprising a source KB and a task KB. The source KB at the transmitter leverages a hierarchical Swin-Transformer, a generative AI scheme, to extract multi-level features from the input image. Concurrently, the counterpart source KB at the receiver utilizes hierarchical residual blocks to generate task-specific knowledge. Furthermore, the task KBs adopt a semantic similarity model to map different task requirements into pre-defined task instructions, thereby facilitating the feature selection of the source KBs. Additionally, we develop a unified residual block-based joint source and channel (JSCC) encoder and two task-specific JSCC decoders to achieve the two image tasks. In particular, a generative diffusion model is adopted to construct the JSCC decoder for the image reconstruction task. Experimental results show that our multi-task generative semantic communication system outperforms previous single-task communication systems in terms of peak signal-to-noise ratio and segmentation accuracy.
\end{abstract}

\section{Introduction}
With the proliferation of artificial intelligence (AI) and the Internet of Things (IoT), there is an increasing demand for communication networks to support a growing number of devices and complex tasks while conserving bandwidth resources. Traditional communication technologies struggle to meet this demand since the total original data needs to be transmitted, resulting in high bandwidth consumption. Semantic communication has been proposed as a promising technology to address this issue \cite{weaver1953recent}. By transmitting the task-relevant semantic information rather than the original data, semantic communication can significantly reduce communication overhead and enhance transmission efficiency \cite{gunduz2022beyond}.

Recently, there have been many studies on semantic communication, which can be categorized into two classes. The first class focuses on utilizing deep neural networks (DNNs) to extract semantic information for transmission \cite{kurka2019successive,kurka2020deepjscc}, while the other class establishes semantic knowledge bases (KBs) to facilitate semantic coding and transmission \cite{yi2023deep,xu2023knowledge,zheng2024generative}. For example, a hierarchical deep learning-based joint source-channel coding (JSCC) scheme was proposed to achieve the successive refinement of image transmission \cite{kurka2019successive}. A deep learning-based JSCC scheme was introduced in \cite{kurka2020deepjscc}, which incorporated channel output feedback to achieve efficient image transmission. Additionally, a semantic communication system based on a shared semantic KB was developed in \cite{yi2023deep}, which combined semantic information with the relevant knowledge from the semantic KB to achieve effective text transmission while reducing communication overhead. 

Although existing studies have achieved great success, they often investigate application scenarios with a single source modality, a specific task requirement, and a particular communication environment. Therefore, the generalization abilities of these designs are often limited \cite{jiang2022reliable}. Some pioneering works have explored multi-task semantic communication \cite{xie2022task,gong2023scalable}. Specifically, the authors of \cite{xie2022task} developed a Transformer-based semantic communication framework to support three tasks, including image retrieval, machine translation, and visual question answering. The authors of \cite{gong2023scalable} proposed a multi-device semantic communication system with multiple channel codecs to enable several image tasks, such as semantic segmentation and object detection. However, these studies need to deploy multiple codecs for different tasks, posing significant challenges for devices with limited storage resources. Moreover, when task requirements change, the systems require retraining, leading to substantial communication and computational overhead.

To address these issues, generative models offer a promising solution due to their powerful self-learning and generalization capabilities. Our previous work \cite{ren2024knowledge} has introduced a generative semantic communication architecture that used generative models to build semantic KBs and enable semantic coding. Building on this, we develop a new multi-task generative semantic communication system in this paper, aiming to support both image reconstruction and segmentation tasks. In our system, both the transmitter and receiver are equipped with a semantic KB, which comprises a source KB and a task KB. The source KB at the transmitter employs a generative model called Swin-Transformer \cite{liu2021swin} to extract multi-level semantic features from the source data, while the counterpart source KB at the receiver utilizes a ResNet \cite{targ2016resnet} to generate task-specific knowledge to facilitate JSCC decoding. Moreover, the task KBs guide the source KBs to output approximate semantic features based on the task requirement, making the system adapt to different image tasks. We evaluate the performance of the proposed system on two image datasets. Experimental results demonstrate that our system outperforms traditional communication systems and classical semantic communication systems in terms of image reconstruction quality and segmentation accuracy.

The rest of this paper is structured as follows. Section II introduces the system model. Section III and Section IV present the specific designs of the semantic KBs and the JSCC codecs, respectively. Section V provides the simulation results and Section VI concludes the paper.
\begin{figure}[t]
    \centering
\includegraphics[width=1\linewidth]{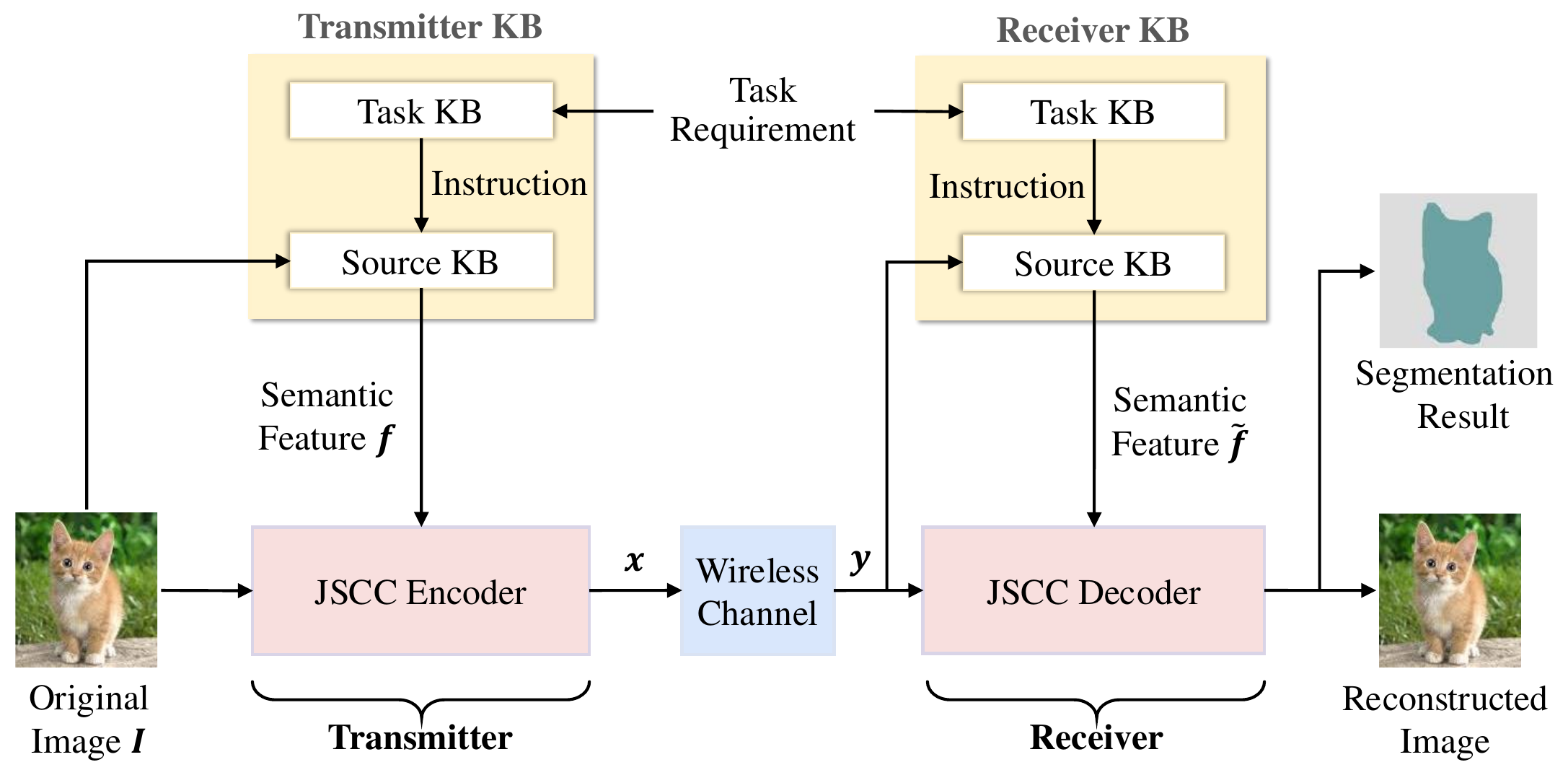}
    \caption{Generative semantic communication system for joint image reconstruction and segmentation.}
    \label{fig:system model}
\end{figure}
\section{Multi-Task Generative Semantic Communication System}
\subsection{System Model}
As shown in Fig. \ref{fig:system model}, we consider a point-to-point semantic communication system, where a transmitter has an input image and a receiver aims to achieve the tasks of image reconstruction and segmentation. To facilitate semantic coding, two semantic KBs are deployed at both the transmitter and receiver. Each semantic KB is composed of a source KB and a task KB. The detailed working mechanism is as follows: 
\begin{itemize}
    \item The task requirement is input into the task KBs at both transmitter and receiver, which align the task requirement with a predefined task list and provide associated task instructions to the source KBs.
    \item The original image $\boldsymbol{I}$ is input into the source KB at the transmitter, which extracts the semantic feature $\boldsymbol{f}$ and delivers it to the JSCC encoder.
    \item The JSCC encoder processes the semantic feature $\boldsymbol{f}$ and the original image  $\boldsymbol{I}$ to obtain the transmitted signal $\boldsymbol{x}$.
    \item The transmitter sends the signal to the receiver through a wireless channel. The received signal is given by
    \begin{equation}
    \boldsymbol{y} = h \boldsymbol{x} + \boldsymbol{n},
    \end{equation}
    where $h$ denotes the channel coefficient, $\boldsymbol{n} \sim \mathcal{CN}(0, \sigma^2 \boldsymbol{U})$ represents an independent and identically distributed circularly symmetric complex Gaussian noise vector with an average noise power of $\sigma^2$, and $\boldsymbol{U}$ is an identity matrix.
    \item The received signal $\boldsymbol{y}$ is input into the receiver's KB and the JSCC decoder, where the source KB at the receiver generates a task-specific feature vector \(\widetilde{\boldsymbol{f}}\) according to the instructions provided by the task KB.
    \item The feature vector \(\widetilde{\boldsymbol{f}}\) is input into the JSCC decoder, which integrates it with the received signal  $\boldsymbol{y}$ to generate the desired result, i.e., the reconstructed image or the segmented image.
\end{itemize} 
\subsection{Task Description} 
To evaluate the proposed system, we consider two typical tasks of image reconstruction and image segmentation, which are described as follows.

\textbf{Image Reconstruction}. 
In this task, the receiver aims to reconstruct the original image. The quality of the reconstructed image is evaluated by a widely-employed metric called peak signal-to-noise ratio (PSNR), which quantifies the ratio between the maximum possible pixel value and the power of the noise and is defined as
\begin{equation}
    \text{PSNR} = 10 \log_{10} \left(\frac{\text{MAX}^2}{\text{MSE}}\right),
\end{equation}
where MAX is the maximum possible pixel value in the image and $\text{MSE}$ denotes the mean squared error between the original image $\boldsymbol{I}$ and the reconstructed image $\hat{\boldsymbol{I}}$. A larger PSNR value indicates a better reconstructed image.

\textbf{Image Segmentation.} In this task, the receiver aims to divide an image into distinct regions, enabling a detailed analysis and understanding of its contents. Each pixel in the image is assigned a label, differentiating between various semantic categories. To optimize the segmentation process, the cross-entropy loss is calculated for each pixel. Moreover, the performance of image segmentation is assessed using the intersection over union (IoU), which is defined as
\begin{equation}
\text{IoU} = \frac{|A \cap B|}{|A \cup B|},
\end{equation}
where $A$ represents the predicted segmentation result and $B$ is the ground-truth result. The numerator $|A \cap B|$ and denominator \(|A \cup B|\) denote the overlap and union between $A$ and $B$, respectively. The IoU ranges from 0 (no overlap) to 1 (complete overlap), measuring the percentage of correctly classified pixels.

\section{Construction of Semantic KBs}
In this section, we provide the specific design of the semantic KBs. We first introduce the two task KBs and then present the source KBs at the transmitter and receiver, respectively.
\subsection{Task KB} 
In this work, we establish two identical task KBs at the transmitter and receiver, which map complex task requirements in natural language into predefined task instructions in the form of discrete tokens. The task KB consists of a memory module and a computing module. The memory module stores multiple predefined task instructions to accommodate various task requirements. In this work, we take image reconstruction and image segmentation as examples. The computing module is used to calculate the semantic similarity score.
Specifically, upon receiving a task requirement, the computing module converts it into vector representations and computes the semantic similarity score between the vectors and each task instruction using the Sentence-BERT model\cite{reimers2019sentence}. The semantic similarity score is defined as 
\begin{equation}
\text{Sim}(r, t_i) = \cos\left( \frac{\boldsymbol{v}(r) \cdot {\boldsymbol{v}}(t_i)}{\| \boldsymbol{v}(r) \| \| \boldsymbol{v}(t_i) \|} \right),
\end{equation}
where $r$ denotes the task requirement, $t_i$ represents the $i$-th task instruction, and $\boldsymbol{v}(r)$ and $\boldsymbol{v}(t_i)$ are their vector representations, respectively. Then, the task KB selects the task instruction with the highest semantic similarity score and delivers it to the source KB for further processing.
\subsection{Source KB at the Transmitter}
\begin{figure}[t]
    \centering
\includegraphics[width=1\linewidth]{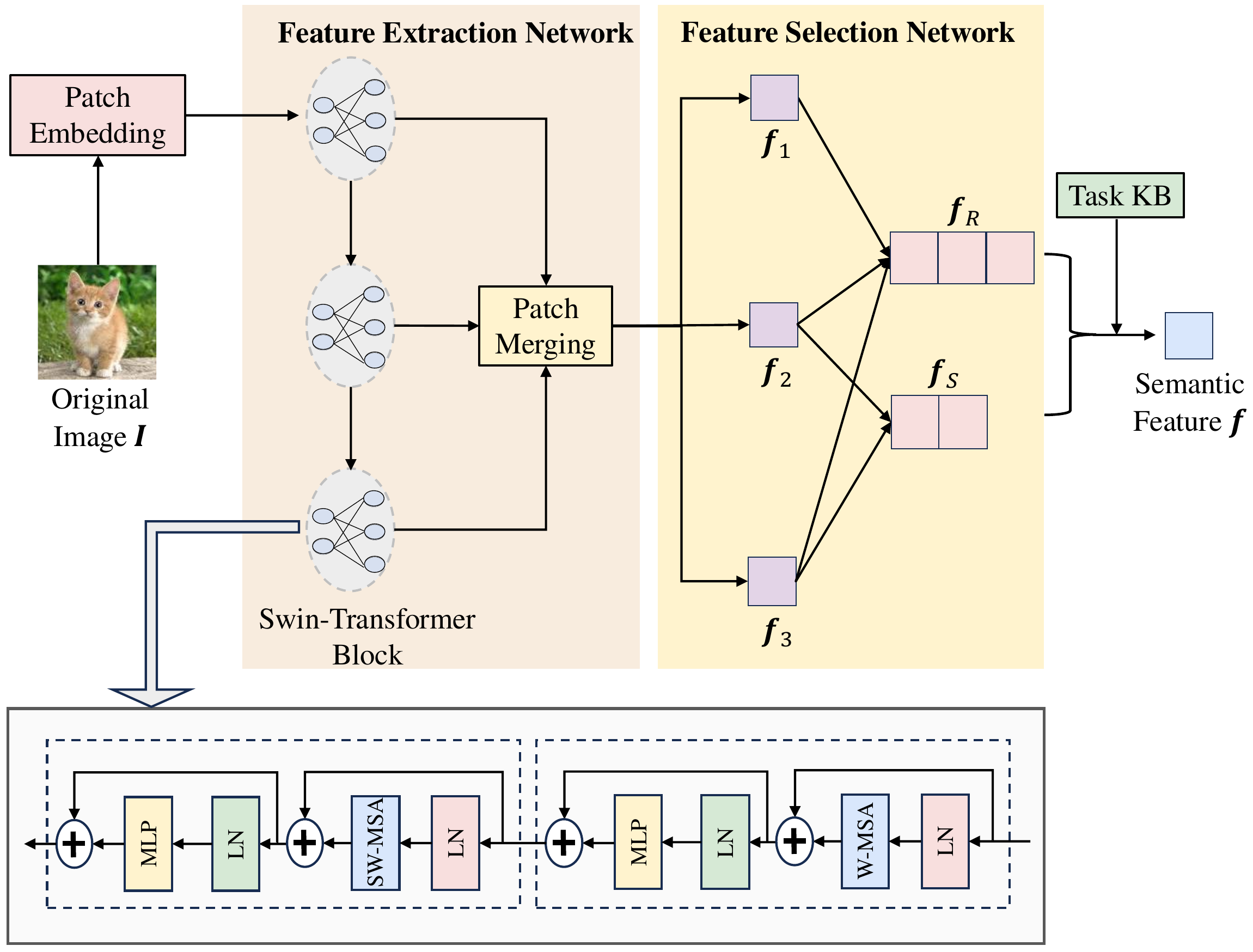}
    \caption{An illustration of the KB at the transmitter.}
    \label{fig:enter-label1}
\end{figure}

The structure of the source KB at the transmitter is shown in Fig. \ref{fig:enter-label1}. It consists of a feature extraction network and a feature selection network.

The feature extraction network is based on a powerful generative model called Swin-Transformer \cite{liu2021swin}, which is shown at the bottom of Fig. \ref{fig:enter-label1}. Specifically, the feature extraction network employs a hierarchical architecture with three Swin-Transformer blocks to generate hierarchical features, capturing semantic information at various levels. In particular, the feature extracted by the first Swin-Transformer block captures the low-level information (e.g., edges and textures). In contrast, the features extracted by the second and third Swin-Transformer blocks capture mid-level information (e.g., local shapes and patterns) and high-level information (e.g., object categories and complex patterns), respectively. The extracted hierarchical features are then processed through patch merging, which compresses spatial information and enhances feature representation, thereby enabling more efficient downstream processing.

Let $\boldsymbol{f}_{1}$, $\boldsymbol{f}_{2}$, and $\boldsymbol{f}_{3}$ denote the low-level, mid-level, and high-level features after patch merging, respectively. These features are then passed through the feature selection network, which selects task-relevant features based on the task instruction provided by the task KB. 
Specifically, since the image reconstruction task requires fine details like edges and colors, we aggregate $\boldsymbol{f}_1$, $\boldsymbol{f}_2$, and $\boldsymbol{f}_3$ to fuse their feature information, thereby improving the reconstruction quality. The aggregated feature is denoted by $\boldsymbol{f}_R = \boldsymbol{f}_1 \oplus \boldsymbol{f}_2 \oplus \boldsymbol{f}_3$, where $\oplus$ denotes the element-wise addition operation. In contrast, the image segmentation task is not sensitive to low-level features. Therefore, we only utilize $\boldsymbol{f}_2$ and $\boldsymbol{f}_3$ to reduce computational cost. The aggregated feature is computed by $\boldsymbol{f}_S = \boldsymbol{f}_2 \oplus \boldsymbol{f}_3$.
\begin{figure}
    \centering
    \includegraphics[width=0.82\linewidth]{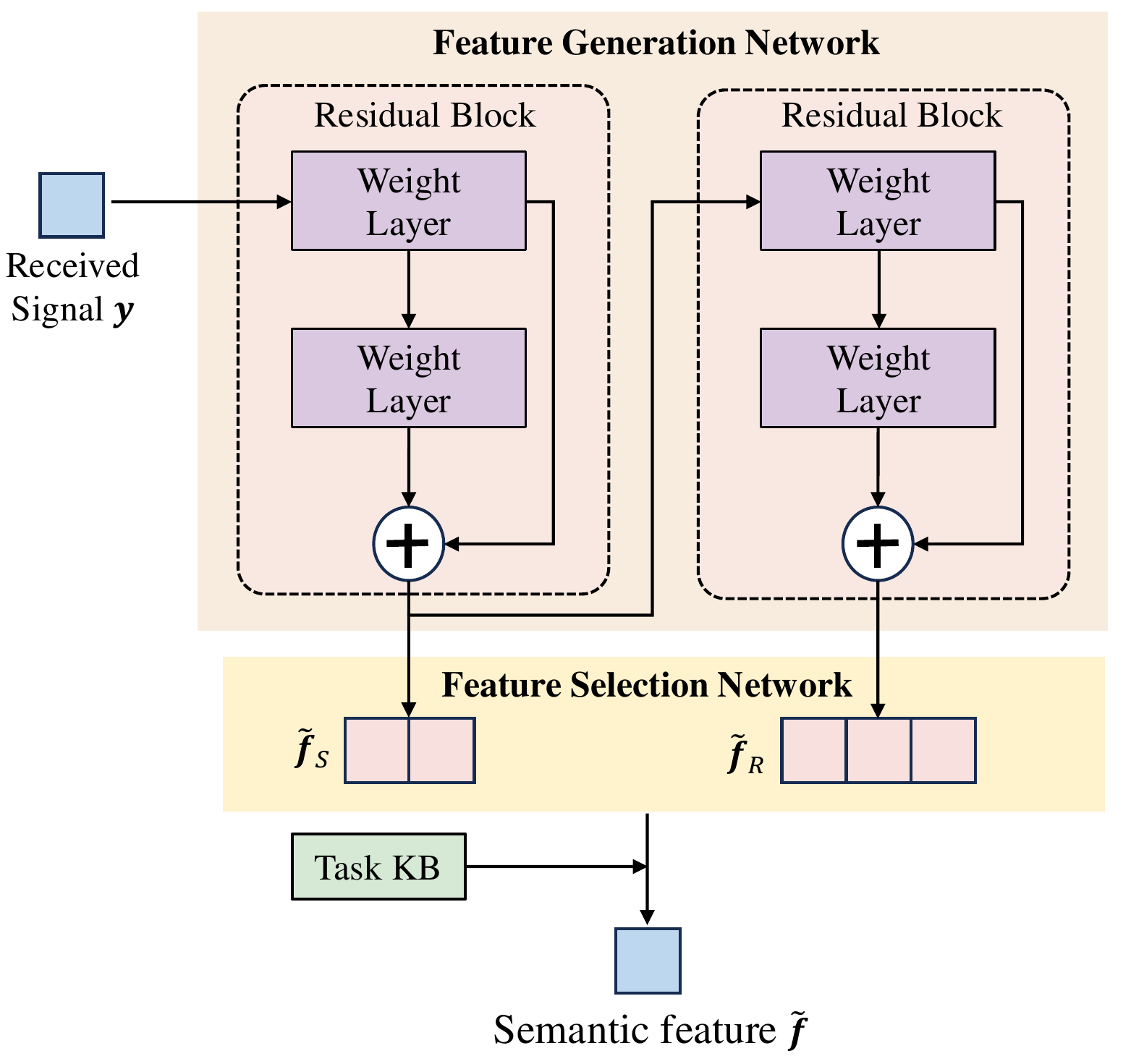}
    \caption{An illustration of the KB at the receiver.}
    \label{fig:enter-label_re}
\end{figure}
\subsection{Source KB at the Receiver} 
As shown in Fig. \ref{fig:enter-label_re}, the source KB at the receiver also consists of two components, including a feature generation network and a feature selection network.

The feature generation network employs two residual blocks that leverage residual connections for efficient feature learning. This architecture progressively generates deep features from the received signal through sequential convolution operations. In each residual block, the noisy signal $\boldsymbol{y}$ is processed through two convolutional neural network (CNN) layers, transforming it into a deep feature. The residual connection provides a direct path for input features, mitigating the vanishing gradient problem during training. Moreover, the residual connection can simplify the learning of feature mappings for each layer, thereby resulting in a more compact feature representation.

The feature selection network is designed to identify task-relevant features from the output of the feature generation network. Specifically, since the image reconstruction task necessitates capturing fine details such as edges and colors, we combine all received features to restore rich image details. Conversely, the image segmentation task prioritizes global features. Therefore, only the high-level feature from the first residual block, i.e., $\widetilde{\boldsymbol{f}}_S$ is utilized. By doing so, the segmentation performance can be guaranteed while the computational cost is reduced.
\section{Joint Source-Channel Encoder and Decoder}
In this section, we present the specific design of the JSCC encoder and decoder to achieve efficient image reconstruction and segmentation.
\subsection{JSCC Encoder}
To reduce the computational and storage costs of the transmitter, we design a unified JSCC encoder for both image reconstruction and segmentation tasks. The JSCC encoder is built upon a ResNet architecture \cite{targ2016resnet}, which incorporates downsampling layers to reduce spatial resolution. As shown in Fig. 1, the input of the JSCC encoder includes two parts, i.e., the original image $\boldsymbol{I}$ and the output feature $\boldsymbol{f}$ from the source KB at the transmitter. The JSCC encoder first concatenates the original image $\boldsymbol{I}$ and the output feature $\boldsymbol{f}$. Then, the concatenated vector is processed by a residual block. To further decrease spatial dimensions and computational complexity, we incorporate downsampling in the JSCC encoder using a specified stride, facilitating multi-scale feature extraction. The final output of the JSCC encoder is the transmitted signal $\boldsymbol{x}$, which is sent to the receiver for image reconstruction or image segmentation.
\subsection{JSCC Decoder}
Due to the distinct objectives of image reconstruction and image segmentation, we develop two independent JSCC decoders at the receiver. As shown in Fig. 1, the input of each JSCC decoder includes two parts, i.e., the received signal $\boldsymbol{y}$ and the semantic feature $\widetilde{\boldsymbol{f}}$ output by the source KB at the receiver. In the following, we introduce the specific design of the two JSCC decoders.

\textbf{JSCC Decoder for Image Reconstruction.}
For image reconstruction, we utilize a generative diffusion model to construct the JSCC decoder due to its powerful generation capabilities \cite{peebles2023scalable}. Specifically, the inputs $\widetilde{\boldsymbol{f}}$  and $\boldsymbol{y}$ are partitioned into a series of patches, which undergo iterative processing with added time step embeddings to capture temporal context. Then, positional encoding is applied to each patch embedding to get spatial information. By doing so, both temporal and spatial information are retained throughout the image reconstruction process. These output embeddings are further refined through a multi-head attention mechanism, capturing the dependencies across different embeddings and enhancing feature representation. After several iterations of attention-based refinement, the results are concatenated to reconstruct the image. In addition, an upsampling layer is utilized to restore the image to its original resolution, followed by a residual block to improve the image quality. 

In the training process, we adopt the MSE between the original image and the reconstructed image as the loss function, which can be expressed as
\begin{equation}
   L_R= \frac{1}{N} \sum_{i=1}^N (p_i - \hat{p}_i)^2,
\end{equation}
where \(N\) is the total number of pixels, \(p_i\) and \( \hat{p}_i \) represent the value of the \(i\)-th pixel in the original and reconstructed images, respectively.

\textbf{JSCC Decoder for Image Segmentation.}
For image segmentation, we utilize a ResNet with two residual blocks to build the JSCC decoder. Each residual block processes the semantic feature $\widetilde{\boldsymbol{f}}$ through convolution operations, with output from each block being added to the original input. This skip connection preserves essential spatial information across layers. Moreover, an upsampling operation is applied to restore the original resolution. Finally, a \(1 \times 1\) convolution operation maps the upsampled features to class logits, obtaining the final segmented result. 

In the training process, we utilize the cross entropy as the loss function, which is defined as
\begin{equation}
    L_S = -\frac{1}{N} \sum_{i=1}^N \log\left( \frac{e^{m_{t_i, q_i}}}{\sum_{j=1}^C e^{m_{j,q_i}}} \right),
\end{equation}
where $t_i$ and $q_i$ are the ground-truth and predicted classes of pixel \(i\), $m_{t_i,q_i}$ is the probability of the predicted class $t_i$ equals the ground-truth class $q_i$, and $C$ is the number of classes.

\section{Experimental Results}
In this section, we conduct experiments to demonstrate the effectiveness of the proposed method. All experiment codes are implemented in a Linux environment equipped with two NVIDIA Tesla A100 40GB GPUs. 
\subsection{Experimental Setup}
\textbf{Datasets.} For the image reconstruction task, we use the DIV2K dataset \cite{agustsson2017ntire} for training and utilize both low-resolution $(224\times224)$ and high-resolution $(768\times512)$ images for validation. For the image segmentation task, we use the PASCAL VOC dataset\cite{everingham2010pascal} for both training and validation. 

\textbf{Baselines.}  For the image reconstruction task, we consider four baseline methods: 1) JPEG+LDPC+QAM(16/64), which adopts conventional separate source and channel coding and utilizes classical QAM for modulation; 2) wireless image transmission transformer (WITT) \cite{yang2023witt}, which uses Swin-Transformer to extract semantic information for transmission; 3) Deep JSCC \cite{bourtsoulatze2019deep}, in which the JSCC encoder consists of five convolutional layers while the JSCC decoder uses transposed convolution layers to reconstruct the image; 4) ADJSCC \cite{xu2021wireless}, which dynamically adjusts the coding strategy to optimize resource utilization, achieving efficient transmission. 

For the image segmentation task, we adopt two baseline methods: 1) JPEG+LDPC+QAM(16/64)+FCN, which uses conventional separate source and channel coding, while utilizing a fully-connected network \cite{he2016deep} to segment images at the receiver; 2) Deeplab JSCC \cite{chen2017deeplab}, which adopts the Deeplab network as the fundamental architecture for the JSCC encoder and decoder. Through atrous convolution, it captures the multi-scale contextual information to achieve effective image segmentation.
\addtolength{\topmargin}{0.03in}
\begin{table}[t]
  \caption{System Parameters}
  \label{tab:model config}
  \centering
  \begin{tabular}{ ccc }
    \toprule
    \textbf{Module} & \textbf{Configuration} & \textbf{Value}  \\
    \midrule
    \multirow{4}{*}{Transmitter KB}          
                 &  Window size   & 7\\ 
                 & Patch size     & 4\\ 
                 & Attention head & 3\\ 
                 & Embedding dim  & 96\\
    \midrule
    \multirow{2}{*}{Receiver KB}
                 & Conv2D layers        & 3\\  
                 & Conv2D kernel size   & 3\\     
    \midrule
    \multirow{3}{*}{JSCC encoder }
                & Conv2D layers         & 3\\  
                & Conv2D kernel size    & 3\\ 
                & Hidden dimension      & 128\\ 
    \midrule
    \multirow{3}{*}{JSCC decoder (reconstruction)}
                & Depth                 & 28\\
                & Patch size            & 2\\ 
                & Attention head        & 16\\
    \midrule
    \multirow{4}{*}{JSCC decoder (segmentation)} 
                & Conv2D layers         & 1\\ 
                & Conv2D kernel size    & 3\\
                & Drop out              &0.1\\ 
                & Num classes           & 21\\ 
    \bottomrule
  \end{tabular}
\end{table}

\textbf{Model Parameters.} 
We perform end-to-end training for all methods. Specifically, we use AdamW as the optimizer with a learning rate of 0.0001 to control the step size during optimization. Momentum coefficients are set to 0.5 and 0.999 to accelerate convergence and reduce oscillations. A weight decay of 0.01 is applied to prevent overfitting. Other model parameters are listed in Table \ref{tab:model config}.
\subsection{Performance on Image Reconstruction}
\begin{figure}[t]
    \centering
\includegraphics[width=1\linewidth]{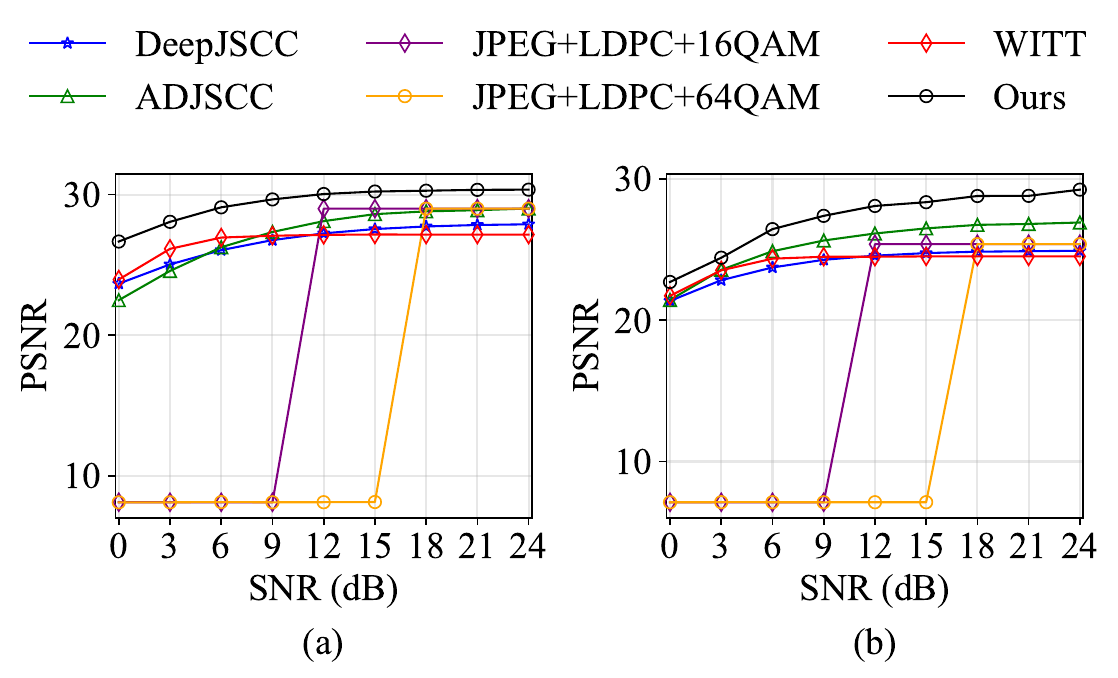}
    \caption{PSNR v.s. SNR in (a) low-resolution image case; (b) high-resolution image case.}
    \label{fig:comparison}
\end{figure}

\begin{figure}[t]
    \centering
    \includegraphics[width=0.99\linewidth]{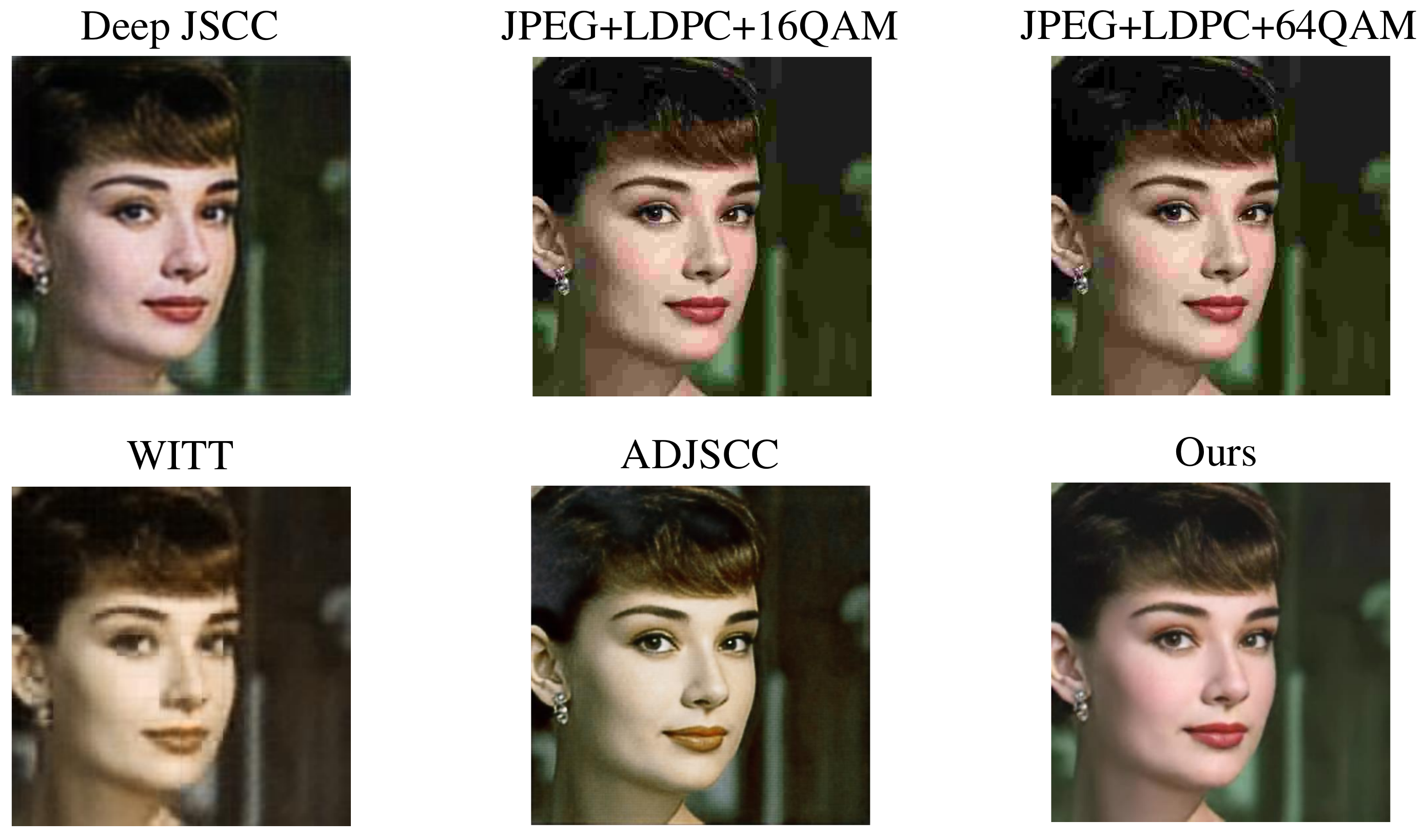}
    \caption{Visualization of reconstructed images (SNR = 18dB).}
    \label{fig:visulization}
\end{figure}

Fig. \ref{fig:comparison} shows the performance comparison of the proposed method with the four baselines for both low-resolution and high-resolution images. In the low-resolution image case, our method consistently achieves the best reconstruction performance among all methods, particularly outperforming the ADJSCC and WITT. This result highlights the importance of semantic KBs in image reconstruction. In the high-resolution image case, our method performs slightly better than other baselines when SNR is low. This is because high-resolution images contain more texture information than low-resolution images. As a result, even minor noise or pixel loss will significantly affect the perceived quality of the reconstructed image. Therefore, the impact of noise is more pronounced on high-resolution images when SNR is low. However, the performance gaps between our method and other baselines increase with SNR, which demonstrates the effectiveness of the proposed method when the communication environment is fine. For a better comparison, we visualize the reconstructed images of different methods when SNR = 18 dB, as shown in Fig. \ref{fig:visulization}. It is observed that our method achieves the best visual quality with smooth facial details. In contrast, the visual results of WITT and ADJSCC exhibit noticeable color deviations. It is because both methods prioritize restoring structural information, such as edges and contours, at the expense of color information preservation. It is worth mentioning that our method also achieves the lowest communication overhead among all methods.
\subsection{Performance on Image Segmentation} 
\begin{figure}[t]
    \centering
    \includegraphics[width=0.99\linewidth]{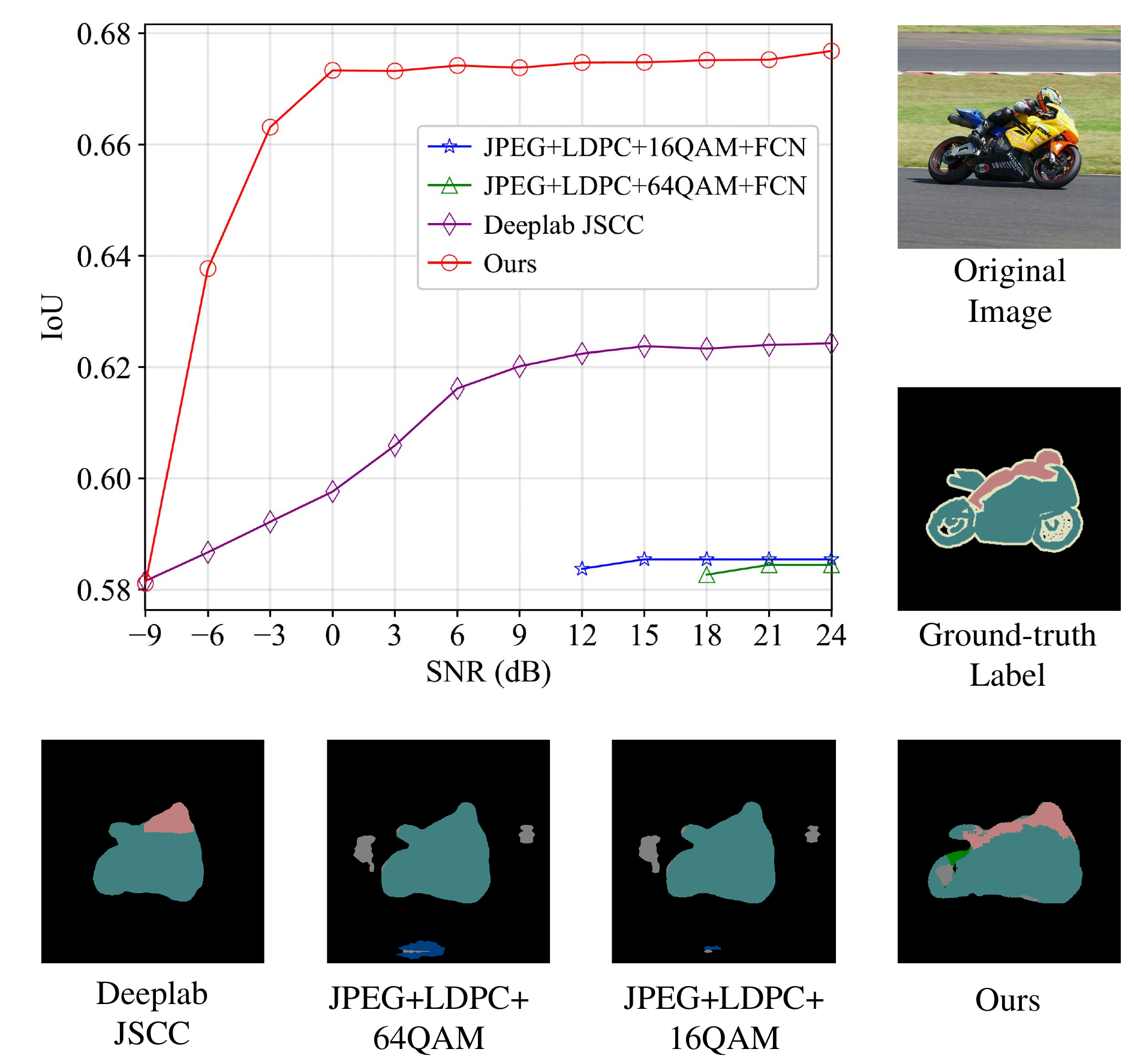}
    \caption{Segmentation accuracy and visualization results of different methods.}
    \label{fig:segmentation}
\end{figure}

Fig. \ref{fig:segmentation} shows the image segmentation performance of all methods. For a fair comparison, the communication overhead of all methods is bounded by 20 Kbits. It is observed that under the same communication overhead, our method always achieves higher image segmentation accuracy compared to other baseline methods. In particular, when the SNR is greater than 0 dB, the performance of our method will stabilize, demonstrating its robustness. Moreover, for traditional communication methods (LDPC+JPEG+QAM(16/64)+FCN), when the SNR is low ($<$12dB for 16QAM or $<$18dB for 64QAM), the receiver cannot reconstruct the original image and the FCN model cannot perform effective image segmentation. Hence, the image segmentation accuracy under low SNR is omitted. The bottom half of Fig. \ref{fig:segmentation} provides the visualization results of all methods. It is seen that the segmentation result of our method is consistent with the ground-truth label, which not only identifies the target objects but also preserves the edge details. In contrast, the Deeplab JSCC cannot accurately distinguish the target objects while the boundaries are blurred. Besides, the traditional communication methods perform even worse with erroneous areas in the segmentation results.

\section{Conclusion}
In this paper, we propose a generative semantic communication system for both image reconstruction and segmentation tasks. Two generative AI schemes of Swin-Transformer and diffusion model are adopted to build the semantic KBs and JSCC decoder, significantly reducing the communication overhead while improving the transmission efficiency. Experimental results show that our method can achieve better performance than several baseline methods. Our designs reveal the potential of using generative AI schemes to handle multiple tasks and provide a new direction for achieving multi-task generative semantic communication. 

\section*{Acknowledgement}
This work was supported in part by NSFC with Grant No. 62293482, the Basic Research Project No. HZQB-KCZYZ-2021067 of Hetao Shenzhen-HK S\&T Cooperation Zone, the Shenzhen Outstanding Talents Training Fund 202002, the Guangdong Research Projects No. 2017ZT07X152 and No. 2019CX01X104, the Guangdong Provincial Key Laboratory of Future Networks of Intelligence (Grant No. 2022B1212010001), and the Shenzhen Key Laboratory of Big Data and Artificial Intelligence (Grant No. ZDSYS201707251409055). 

\bibliographystyle{IEEEtran}  
\bibliography{references} 
\end{document}